# Chemical pathways of SO$_2$ with hydrogen atoms on interstellar ice analogues

Thanh Nguyen,[1] Yasuhiro Oba,[1] W. M. C. Sameera,[1,2] Kenji Furuya[3,] and Naoki Watanabe[1]

[1]*Institute of Low Temperature Science, Hokkaido University, N19W8, Kita-Ku, Sapporo, Hokkaido, 060-0819, Japan*

[2]*Department of Chemistry and Molecular Biology, University of Gothenburg, SE-412 96 Gothenburg, Sweden*

[3]*Department of Astronomy, Graduate School of Science, University of Tokyo, Tokyo 113-0033, Japan*

Abstract

Sulfur dioxide (SO$_2$) is a sulfur-containing molecule expected to exist as a solid in the interstellar medium (ISM). In this study, we performed laboratory experiments and computational analyses on the surface reactions of solid SO$_2$ with hydrogen atoms on amorphous solid water (ASW) at low temperatures. After 40 min of exposure of SO$_2$ deposited on ASW to H atoms, approximately 80% of the solid SO$_2$ was lost from the substrate at 10–40 K, and approximately 50% even at 60 K, without any definite detection of reaction products. Quantum chemical calculations suggest that H atoms preferentially add to the S atom of solid SO$_2$, forming the HSO$_2$ radical. Further reactions of the HSO$_2$ radical with H atoms result in the formation of several S-bearing species, including HS(O)OH, the S(O)OH radical, HO–S–OH, HS–OH, and H$_2$S. In codeposition experiments involving H and SO$_2$, we confirmed the formation of H$_2$S, HS(O)OH, and/or HO–S–OH. However, the yields of these S-bearing species were insufficient to account for the complete loss of the initial SO$_2$ reactant. These findings suggest that some products desorbed into the gas phase upon formation. This study indicates that a portion of SO$_2$ in ice mantles may remain unreacted, avoiding hydrogenation, while the remainder is converted into other species, some of which may be subject to chemical desorption.

## 1. INTRODUCTION

Sulfur is a crucial element, potentially playing a key role in the chemical evolution leading to the origin of life in space. Approximately 30 S-bearing species have been detected, primarily in the gas phase of the interstellar medium (ISM). However, their observed total abundance remains considerably lower than the expected cosmic abundance of S (Wakelam et al. 2004; Anderson et al. 2013; Fuente et al. 2023). Consequently, astrochemical models frequently propose that the so-called "missing sulfur" may be present in solid form on interstellar grains (Ruffle et al. 1999; Vidal et al. 2017; Laas & Caselli 2019). However, the detection of S-bearing species in the solid state in the ISM has been quite limited. Carbonyl sulfide (OCS) and sulfur dioxide ($SO_2$) have been classified as either "likely identified" or "possibly identified" in the solid state (Boogert et al. 2015). Recently, McClure et al. (2023) distinctly detected solid OCS in molecular clouds toward the background stars NIR38 and J110621 using the James Webb Space Telescope (JWST). Additionally, Rocha et al. (2024) reported the detection of solid $SO_2$ toward protostars NGC 1333 IRAS 2A and IRAS 23385+6053. These highly sensitive astronomical observations may pave the way for the detection of additional solid-state S-bearing species. Atomic S and its allotropes, such as $S_8$, have been proposed as potential candidates for missing S in recent modeling studies (e.g., Shingledecker et al. 2020). However, detecting these S-bearing species with radio telescopes may be challenging owing to their lack of a dipole moment.

While the detection of solid S-bearing species in the ISM remains limited, recent experimental and computational studies have considerably advanced our understanding of the chemical pathways of these species on interstellar icy grains. For instance, hydrogen sulfide ($H_2S$), which is likely formed on dust grains through the successive addition of H atoms to S atoms (Millar et al. 1986; Millar & Herbst 1990), can desorb into the gas phase via chemical desorption with an efficiency of up to 3% (Oba et al. 2018, 2019; Furuya et al. 2022). This process likely contributes to the presence of $H_2S$ in the gas phase. Additionally, more recent studies have reported the formation of disulfane ($H_2S_2$) through surface reactions of H atoms with thick $H_2S$ layers (Santos et al. 2023). Methyl mercaptan ($CH_3SH$), one of the simplest S-bearing organic molecules, was believed to form on interstellar grains through the successive addition of H atoms to carbon monosulfide (CS) (Majumdar et al. 2016; Müller et al. 2016; Vidal et al. 2017; Lamberts 2018). Given that $CH_3SH$ has been detected in the gas phase of the ISM (Majumdar et al. 2016; Müller et al. 2016; Bouscasse et al. 2022), it was expected to desorb upon formation via chemical desorption, similar to $H_2S$. However, our laboratory experiments revealed that chemical desorption is inefficient for $CH_3SH$ (Nguyen et al. 2023). Instead, solid $CH_3SH$ readily dissociates into $CH_3$ and $H_2S$ after reacting with H atoms on icy dust grains, which may explain the lack of solid-state $CH_3SH$ detection (Nguyen et al. 2023). Regarding OCS, it reacts with H atoms on dust grains via quantum tunneling, leading to the formation of thioformic acid (HC(O)SH) (Molpeceres et al. 2021, 2022; Nguyen et al. 2021), which was recently detected in the quiescent cloud G+0.693-0.027 near the Galactic center (Rodríguez-Almeida et al. 2021). HC(O)SH can also form through the reaction of HCO with SH radicals on interstellar dust grains at very low temperatures (Nguyen et al. 2021). Additionally, studies have investigated solid S-bearing species as targets for energetic processes induced by ultraviolet photons and cosmic-ray analogs (e.g., Jimenez-Escobar et al. 2014; Cruz-Diaz et al. 2014; Chen et al. 2015; Mifsud et al. 2021, 2022; Wang et al. 2022; Cazaux et al. 2022; Li et al. 2022), which have greatly enhanced our understanding of sulfur chemistry on interstellar icy grains at extremely low temperatures.

In this study, we focus on the chemical pathways of solid $SO_2$ on icy surfaces. $SO_2$ is not only prevalent in the ISM but also in the atmospheres of planetary bodies such as Venus (Barker 1979; Finlayson-Pitts 2000; Calmonte et al. 2016) and Neptune (Dyrek et al. 2024), making it a key molecule that may provide crucial insights into the chemical evolution of S-bearing species in space. In the ISM, $SO_2$ has a relatively high abundance among S-bearing interstellar molecules (Charnley 1997). Furthermore, $SO_2$ is likely to be present as a solid on interstellar dust grains (Boogert et al. 1997; Zasowski et al. 2009; McClure et al. 2023; Rocha et al. 2024). The estimated abundance of solid $SO_2$ is up to 1% relative to $H_2O$, meaning that solid $SO_2$ sequesters approximately 6% of the cosmic sulfur abundance (Boogert et al. 1997). Therefore, understanding the chemistry of $SO_2$ on icy dust grains could be crucial for interpreting sulfur chemistry in the ISM. However, the reactivity of solid $SO_2$ on icy grains remains poorly understood, highlighting the need for further experimental, theoretical, and observational studies.

Previous experimental and theoretical studies suggest that $SO_2$ forms through various reactions in both hot (>100 K) core regions and cold (~10 K) dense clouds (Cohen & Gross 1969; Millar & Herbst 1990; Moore et al. 2007; Jimenez-Escobar & Muñoz Caro 2011; Chen et al. 2015) as follows:

$$SO + O \rightarrow SO_2 + h\nu, \quad (1)$$

$$SO + OH \rightarrow SO_2 + H, \quad (2)$$

$$SO + SO \rightarrow SO_2 + S. \quad (3)$$

Given the potential presence of solid $SO_2$ in the ISM, it is likely to interact with H atoms, which are the predominant reactants on interstellar icy dust (Hama & Watanabe 2013). While the reaction of $SO_2$ with H atoms has been studied experimentally in the gas phase at near and above room temperatures (Morris et al. 1995; Blitz et al. 2006) and in solid para-$H_2$ (Gobi et al. 2021), the chemistry of solid $SO_2$ under astrophysically relevant conditions remains unclear. Further systematic investigations into possible reaction pathways are needed to illuminate the role of $SO_2$ in dense clouds. We performed laboratory experiments and computational studies on the surface reaction of solid $SO_2$ with H atoms at low temperatures (10–60 K) to gain a deeper understanding of the physico-chemical behavior of $SO_2$ and the potential S-bearing products under conditions that simulate those in the ISM.

## 2. EXPERIMENTAL PROCEDURE

All experiments were performed using a system known as the Apparatus for Surface Reaction in Astrophysics (ASURA). Details of the ASURA system can be found in previous studies (Watanabe et al. 2006; Nagaoka et al. 2007; Nguyen et al. 2020, 2021, 2023). The system primarily consists of an ultrahigh vacuum chamber with a base pressure of $10^{-8}$ Pa, an aluminum substrate (Al) mounted on a He cryostat, an atomic source, a Fourier-transform infrared spectrometer (FTIR), and a quadrupole mass spectrometer (QMS). The surface temperature can be accurately controlled from 5 to 300 K.

We investigated the chemical reactions of solid $SO_2$ with H atoms on a compact amorphous solid water (c-ASW) substrate at various temperatures. c-ASW was prepared by vapor depositing water onto a surface held at 110 K. The resulting c-ASW layer thickness was estimated to be 20 monolayers (MLs; 1 ML = $1 \times 10^{15}$ molecules $cm^{-2}$). Surface reactions were performed on c-ASW at temperatures of 10, 30, 40, or 60 K.

Gaseous $SO_2$ was predeposited onto the c-ASW substrate at a rate of 1 ML minute$^{-1}$, resulting in an equivalent abundance of approximately 1 ML. This was determined using the peak area of the $SO_2$ stretching band at 1347 cm$^{-1}$ and the band strength of $1.5 \times 10^{-17}$ cm molecule$^{-1}$ (Garozzo et al. 2008). In the present study, we estimated the column density of molecules in reflection absorption spectroscopy using transmission band strengths, which could cause some potential errors in the derived column densities (e.g. Ioppolo et al. 2013). However, we have already confirmed that the difference in the column density was within a factor of two when we used transmission band strengths for estimating the column density of molecules in reflection modes under the current setup (Oba et al. 2009). Hence, we are confident that such potential errors would not significantly alter the present results and their interpretation. H (or D) atoms were generated by dissociating $H_2$ (or $D_2$) in a microwave-discharged plasma in the atomic source. These H (D) atoms were then cooled to 100 K through multiple collisions with the inner wall of the Al pipe, which was maintained at 100 K (Nagaoka et al. 2007). Using the method outlined by Oba et al. (2014), the flux of H and D atoms was estimated to be $5.0 \times 10^{14}$ and $3.4 \times 10^{14}$ cm$^{-2}$ s$^{-1}$, respectively. Predeposited solid $SO_2$ was exposed to H or D atoms for up to 2 h for the measurement of the reaction kinetics.

We performed additional experiments to investigate the codeposition of $SO_2$ and H atoms on the Al substrate at 10 K to more accurately identify potential reaction products. The flux of H (D) atoms was the same as in the predeposition experiment, and the deposition rate of $SO_2$ was approximately 0.3 ML minute$^{-1}$.

The interaction between $SO_2$ and H (D) atoms was monitored in situ using FTIR with a resolution of 2 cm$^{-2}$. Reactants and products desorbed from the substrate were detected with the QMS through thermal programmed desorption (TPD) at a ramping rate of 4 K minute$^{-1}$.

## 3. COMPUTATIONAL METHODS

Structure optimizations were performed using the ωB97X-D functional (Chai & Head-Gordon 2008) and the def2-TZVP basis set (Weigend & Ahlrichs 2005), as implemented in the Gaussian 16 program (Frisch et al. 2016). Vibrational frequency calculations were performed to confirm the nature of the optimized structures (i.e., no imaginary frequencies for local minima and one imaginary frequency for transition states) and to calculate zero-point energies. Ice cluster models (Sameera et al. 2021), with the ASW cluster model consisting of 60 $H_2O$ molecules and the $I_h$ cluster model consisting of 48 $H_2O$ molecules, were used to calculate the binding energies of $SO_2$ on ice (Fig. 1). During structure optimizations, the outermost $H_2O$ molecules on the sides and bottom of the cluster were held fixed (Fig. 1, marked in green). The reaction mechanism between $SO_2$ and H on the ice was calculated using the ASW ice cluster, with all $H_2O$ molecules in the cluster held fixed during the reaction path search.

## 4. RESULTS AND DISCUSSION

### 4.1. *Reactions of predeposited $SO_2$ with H or D atoms on c-ASW at different temperatures*

Figure 2 shows the FTIR spectrum of solid $SO_2$ on c-ASW at 10 K. Two prominent absorption bands are observed at 1347 and 1154 cm$^{-1}$. The peak shapes and positions align well with those reported in previous studies, which attribute these bands to the antisymmetric stretching vibrational mode ($v_3$) and the symmetric stretching band ($v_1$), respectively (Nxumalo & Ford 1995; Garozzo et al. 2008; Mifsud et al. 2023). Additionally, a minor infrared band appeared at 2460 cm$^{-1}$, which is attributed to the $v_1 + v_3$ combination band of solid $SO_2$ (Moore et al. 2007; Loeffler & Hudson 2010; Mifsud et al. 2023).

Figure 3a shows the difference in the spectra of solid $SO_2$ on c-ASW at 10 K after exposure to H atoms compared to the initial spectrum. Peaks below the baseline, which indicate the depletion of the initial reactant, were observed at 1347 and 1154 cm$^{-1}$ after exposure to H atoms. Because solid $SO_2$ remained unchanged after interactions with $H_2$ molecules, the observed decrease in $SO_2$ is attributed to its reaction with H atoms on the substrate. The following two reactions are potential initial steps for the reaction of H atoms with solid $SO_2$:

$$SO_2 + H \rightarrow HOSO, \qquad (4)$$

$$SO_2 + H \rightarrow HSO_2, \qquad (5)$$

We found that reaction (4), involving H addition to the O atom of $SO_2$, has a high activation barrier ($E_a$ = 0.40 eV or 4642 K; Fig. 4). In contrast, the activation barrier for reaction (5), where H adds to the S atom ($E_a$ = 0.07 eV or 812 K; Fig. 4), is considerably lower. This strongly indicates that solid $SO_2$ is primarily consumed through reaction (5). The resulting $HSO_2$ radical can react with an additional H atom, which requires no activation barrier, to form sulfinic acid, HS(O)OH:

$$HSO_2 + H \rightarrow HS(O)OH, \qquad (6)$$

However, we did not observe any distinct infrared features corresponding to the potential products ($HSO_2$ and HS(O)OH) in the $SO_2$ predeposition experiment (Fig. 3a). The H abstraction from $HSO_2$ ($HSO_2 + H \rightarrow SO_2 + H_2$) has a substantial activation barrier (0.40 eV or 4642 K), making this backward reaction unlikely to occur. Even if it does, the product $SO_2$ would react with another H atom, regenerating $HSO_2$. Our quantum chemical calculations suggest that the formed HS(O)OH could further react with H atoms as follows:

$$HS(O)OH + H \rightarrow H_2S(O)OH, \qquad (7)$$

$$HS(O)OH + H \rightarrow OSOH + H_2. \qquad (8)$$

Reaction (7) has a relatively high activation barrier (0.23 eV or 2669 K; Fig. 4), whereas reaction (8) has a lower activation barrier (0.08 eV or 928 K). Therefore, reaction (8) is more likely to occur than reaction (7). Subsequently, the product of reaction (8), the OSOH radical, reacts with an H atom to form sulfoxylic acid (reaction 9), HOSOH (Fig. 4):

$$OSOH + H \rightarrow HOSOH. \qquad (9)$$

Further reactions of HOSOH with H atoms proceed through the following pathways:

$$HOSOH + H \rightarrow HO \cdots S(H)OH, \qquad (10)$$

$$\text{HO}\cdots\text{S(H)OH} + \text{H} \rightarrow \text{HSOH} + \text{H}_2\text{O}, \quad (11)$$

$$\text{HSOH} + \text{H} \rightarrow \text{H}_2\text{S} + \text{OH}, \quad (12)$$

$$\text{OH} + \text{H} \rightarrow \text{H}_2\text{O}. \quad (13)$$

Reactions (10) and (12) have relatively low activation barriers of 0.13 eV (1509 K) and 0.09 eV (1044 K), respectively, while other reactions are barrierless. Despite this, no S-bearing products were detected by FTIR analysis, as noted earlier. One possible explanation for this is that the products of reactions (5)–(12) may have desorbed from the surface after formation, potentially via chemical desorption. The current experimental setup did not allow for the detection of these desorbed species during the H atom exposure period. Additionally, the low abundance of the initial reactant (~1 ML of $SO_2$) on the surface could also contribute to the difficulty in detecting these products.

Figure 3b depicts the change in the relative abundance of solid $SO_2$ after exposure to H atoms at temperatures of 10, 30, 40, and 60 K with respect to atom exposure times. After up to 2 h of exposure, the fractional loss of $SO_2$ was approximately 80% at 10–40 K and 40% at 60 K, compared to the initial amount of solid $SO_2$. This loss is assumed to be due solely to the formation of $HSO_2$ and its subsequent chemical desorption. The desorption efficiency of $HSO_2$ per incident atom was roughly estimated from the slope of the plot in Fig. 3 to be 1% at 10–40 K and 0.2% at 60 K. These values are comparable to those estimated for solid $H_2S$ using the same method (Oba et al. 2018). This desorption efficiency should be considered an upper limit because it is certain that a portion of $HSO_2$ is consumed as a reactant in reaction (6), as demonstrated in the codeposition experiment (see section 4.2).

Because the loss of $SO_2$ is caused by the reaction with H atoms only, as mentioned above, the variation in the relative abundance of $SO_2$ can be represented by the following rate equation:

$$d[\text{SO}_2]/dt = -k_\text{H}[\text{H}][\text{SO}_2], \quad (14)$$

where $[SO_2]$ and $[H]$ are the surface number densities of $SO_2$ and H atoms on the ice surface, respectively, and $k_H$ is the rate coefficient for the H addition reactions. The reaction kinetics can be derived using the following integrated equation (15):

$$\Delta[\text{SO}_2]_t/[\text{SO}_2]_0 = A \times (1 - \exp(-k'_\text{H} t)), \quad (15)$$

where $\Delta[SO_2]_t$ is the change in the abundance of solid $SO_2$ at time $t$, and $[SO_2]_0$ denotes the initial $SO_2$ abundance. $A$ is the saturation value for the decrease in $SO_2$, and $k'_H$ ($= k_H[H]$) is the effective rate constant for the reaction of $SO_2$ with H atoms. By fitting the data shown in Fig. 3b to equation (15), we determined $k'_H$ values of $(2.2 \pm 0.2) \times 10^{-3}$, $(1.7 \pm 0.1) \times 10^{-3}$, $(2.4 \pm 0.1) \times 10^{-3}$, and $(0.8 \pm 0.1) \times 10^{-3}$ s$^{-1}$ at 10, 30, 40, and 60 K, respectively. Except at 60 K, the $k'_H$ values are nearly independent of the surface temperature, which contrasts with the behavior observed in other surface reactions involving H atoms, where $k'_H$ decreases with increasing surface temperature (Watanabe & Kouchi 2008; Hama & Watanabe 2013; Oba et al. 2016, 2019; Nguyen et al. 2020, 2023). Notably, solid SO$_2$ loss occurred even at 60 K, despite the extremely short residence time of H atoms on c-ASW (<10$^{-8}$ s, assuming a binding energy of 650 K for H atoms on ices, Al-Halabi & van Dishoeck 2007). This indicates that the reaction predominantly proceeded via the Eley–Rideal (ER) mechanism at 60 K. In contrast, at temperatures below 40 K, both the ER and Langmuir–Hinshelwood (LH) mechanisms may contribute to the consumption of solid SO$_2$, resulting in significant loss from the icy surfaces.

When H atoms were substituted with D atoms in the predeposition experiment, a decrease in SO$_2$ was observed (spectra not shown), similar to the results obtained with H atoms (Fig. 3a). No S-bearing species were identified in either case. These findings suggest that reaction (5) occurs very rapidly even with D atoms, and some of the reaction products may desorb immediately upon formation. Figure 5 shows the variations in the relative abundance of SO$_2$ with D atoms on c-ASW at 10 K, compared to the results with H atoms. The data in Fig. 5 were fitted using equation (15). We determined the effective rate constant $k'_D$ (= $k_D$[D], where $k_D$ is the rate constant for the addition of D to SO$_2$, and [D] is the surface number density of D atoms) to be $4.4 \pm 0.3 \times 10^{-3}$ s$^{-1}$. This value is twice as large as $k'_H$ at the same temperature (10 K), indicating that the loss of SO$_2$ occurs more rapidly with D atoms than with H atoms under the given experimental conditions. However, this does not necessarily imply that the reaction rate constant $k_D$ is greater than $k_H$ because the effective rate constant $k'_X$ (where X = H or D) is the product of $k_X$ and [X]. According to Kuwahata et al. (2015), the number density ratio of D to H atoms ([D]/[H]) is approximately 4 on porous (p)-ASW and approximately 10 on polycrystalline H$_2$O under the current atom fluxes. Assuming that the [D]/[H] ratio on c-ASW is between 4 and 10, for example, 7, the ratio $k'_D/k'_H = k_D[D]/k_H[H] = 2$ can be rearranged as follows:

$$k_D/k_H = 2/7 = 0.28. \qquad (16)$$

A $k_D/k_H$ value of 0.28 means that the D addition is slower than the H addition, which may be reasonable as an isotope effect on reaction (5) with a small barrier of 0.07 eV (Fig. 4).

### 4.2. Identification of products after codeposition of $SO_2$ and H or D atoms at 10 K

Figure 6a presents the variations in the FTIR spectrum of solid $SO_2$ after a 2-h codeposition with H atoms on the Al substrate at 10 K, compared to the spectrum obtained from a codeposition with $H_2$ molecules (blank) for reference. In the blank experiment, no additional peaks were observed beyond those from solid $SO_2$. In contrast, after codeposition with H atoms, several new peaks appeared that were not observed in the predeposition experiment (Fig. 3). The most prominent peak appeared at approximately 3400 $cm^{-1}$, accompanied by a broad absorption centered at 1676 $cm^{-1}$. These features are consistent with those of amorphous $H_2O$ (Oba et al. 2009). Because $H_2O$ was not produced in the blank experiment (black trace in Fig. 6a), its formation must be attributed to the reactions between solid $SO_2$ and H atoms. Although the formation of water was inferred from the OH stretching band region in the predeposition experiment (Fig. 3(a)), it was not clearly visible owing to the interference from the predeposited c-ASW. Additionally, $H_2S$ was detected, as indicated by the S–H stretching band at 2559 $cm^{-1}$. The column densities of $H_2O$ and $H_2S$ were estimated from the peak areas in Fig. 6(a) and their band strengths ($2.0 \times 10^{-16}$ cm molecules$^{-1}$ for the O–H stretching band of $H_2O$ and $8.3 \times 10^{-18}$ cm molecules$^{-1}$ for the S–H stretching band of $H_2S$), (Gerakines et al. 1995; Fathe et al. 2006) to be approximately 8.4 and 2.7 ML, respectively. Given that the total consumption of $SO_2$ was approximately 23 ML in the codeposition experiment—estimated from the difference in $SO_2$ abundances between the $SO_2 + H_2$ (~29 ML) and $SO_2 + H$ experiments (~6 ML) after 2 h—the yields of $H_2O$ and $H_2S$ were approximately 18% and 12%, respectively. Because $H_2S$ could have been partially lost from the substrate owing to chemical desorption (Oba et al. 2018), this value should be considered a minimum estimate. This observation does not conflict with the non-detection of $H_2S$ in the predeposition experiment reported previously (Fig. 4a).

One potential S-bearing product that may be present on the substrate is HOSOH (reaction 9). Gobi et al. (2021) demonstrated that HOSOH exhibited an absorption at approximately 790 $cm^{-1}$ in a para-$H_2$ matrix. If the broad peak observed at approximately 800 $cm^{-1}$ in Fig. 6 is solely attributed to HOSOH, its column density was roughly estimated to be 4.6 ML, based on the peak area and the band strength for this peak ($2.8 \times 10^{-17}$ cm molecule$^{-1}$; Gobi et al. 2021). However, because the observed peak may also include contributions from other S-bearing species such as HSOH and HS(O)OH (Gobi et al. 2021; Smardzewski & Lin 1977), this estimate should be considered tentative. Nevertheless, it is clear that the abundance of HOSOH and other S-bearing species is considerably lower than the amount of $SO_2$ consumed in the codeposition experiment. Additionally, radical products such as $HSO_2$ and S(O)OH likely form as

10intermediates in the reaction sequence (Fig. 4). Owing to their high reactivity, these intermediates are expected to be quickly consumed by reactions with H atoms, making their detection challenging under the current experimental conditions. Accurate assignment of the observed IR peaks and quantification of new products were not feasible, primarily owing to the lack of reference data for the band positions of potential products in the solid state. In addition, during warming up, some ionic S-bearing species such as $S_2O_5^{2-}$ and $HSO_3^-$ may form by reactions of $SO_2$ with $H_2O$ (Loeffler & Hudson 2010; Kaňuchová et al. 2017), which could have contributed to the newly appeared peaks at 971 cm$^{-1}$ at >130 K (Fig. 6b).

Figure 7a shows the results of the TPD experiment at $m/z$ = 34 after the codeposition of $SO_2$ and H atoms or $H_2$ for 2 h at 10 K. The peaks observed at approximately 80 K and 155 K in panel (a) correspond to the desorption of $H_2S$ from $H_2S$ solids and from $H_2O$-trapped $H_2S$, respectively (Collings et al. 2004). A smaller peak at 110 K in the blank experiment is attributed to the fragment ($^{34}S$) from $^{34}SO_2$. Therefore, the production of $H_2S$ was confirmed in the case of H atoms by the desorption peak at $m/z$ = 34 in the TPD experiment (Fig. 7a).

The TPD experiment provided insights into the presence of HS(O)OH and/or HOSOH after codeposition. Figure 7b shows the TPD results at $m/z$ = 66, which corresponds to the mass of HS(O)OH and HOSOH after a 2-h codeposition of $SO_2$ with H atoms (solid line) or $H_2$ molecules (blank: dotted line). In the $SO_2$ + H experiment, a desorption peak was observed at approximately 160 K. In contrast, the blank experiment showed only a single peak at approximately 110 K. Based on the desorption temperature (~110 K for $SO_2$, Collings et al. 2004) and the peak area, which was approximately 4% of $^{32}S^{16}O_2$ at $m/z$ = 64, the peak at approximately 110 K can be attributed to one of the isotopologues of $SO_2$, specifically $^{34}S^{16}O_2$ (with $^{34}S$ naturally occurring at approximately 4% of $^{32}S$, as noted in the NIST database; hereafter, elements are referred to by their most abundant isotopes, e.g., "S" refers to "$^{32}S$"). The desorption peak area at ~160 K in the $SO_2$ + H experiment ($m/z$ = 66, Fig. 7b) was about 5% of that for $SO_2$ ($m/z$ = 64) in the same experiment. Based on the desorption temperature of $SO_2$ in the presence of water (~150 K, Collings et al. 2004) and the natural abundance of $^{34}S$, $^{34}SO_2$ may have a large contribution to the observed peak at ~160 K in Fig. 7(b). In addition, we expect this peak could also have contributions from other S-bearing molecules such as HS(O)OH and HOSOH, which have a parent mass of 66. This conclusion is consistent with our assumption that the observed IR peak at ~1200 cm$^{-1}$ is attributed to HS(O)OH and HOSOH because this peak disappeared upon warming to 160–170 K (Fig. 6b).

After codepositing $SO_2$ with D atoms on the Al substrate at 10 K, we observed several absorption bands not seen in the blank experiment (i.e., the codeposition of $SO_2$ with $D_2$ molecules) (Fig. 8). This suggests that these peaks originate from products formed by the possible reactions detailed below.

$$SO_2 + D \rightarrow DSO_2, \quad (17)$$

$$DSO_2 + D \rightarrow DS(O)OD, \quad (18)$$

$$DS(O)OD + D \rightarrow S(O)OD + D_2, \quad (19)$$

$$S(O)OD + D \rightarrow DOSOD, \quad (20)$$

$$DOSOD + D \rightarrow DO\cdots S(D)OD, \quad (21)$$

$$DO\cdots S(D)OD + D \rightarrow DSOD + D_2O, \quad (22)$$

$$DSOD + D \rightarrow D_2S + OD, \quad (23)$$

$$OD + D \rightarrow D_2O. \quad (24)$$

Reactions (18), (20), (22), and (24) are barrierless (see section 5 for details). The prominent peak at 2540 cm$^{-1}$, along with the broad peak at 1220 cm$^{-1}$, can reasonably be attributed to the O–D stretching and bending bands of solid $D_2O$, respectively (Watanabe et al. 2000). The presence of $D_2O$ in the product is further supported by the detection of a desorption peak at m/z = 20 during the TPD experiment (Fig. 9a). The column density of the formed $D_2O$ during codeposition was estimated from the peak area and the band strength of the O–D stretching band (Oba et al. 2014) to be approximately 3.5 ML. This represents approximately 15% of the $SO_2$ consumption in the same experiment (22.5 ML). Assuming that two $D_2O$ molecules are produced from one $SO_2$ molecule through a series of $SO_2$ + D reactions (reactions 17–24), the estimated yield of $D_2O$ is approximately 8%, which is lower than the $H_2O$ yield of approximately 18%. This lower yield is likely attributed to the isotope effect on reactions with an activation barrier (e.g., reaction 21), where D-addition typically occurs more slowly than H addition (Hama & Watanabe 2013). Additionally, a small absorption observed at 1860 cm$^{-1}$ can be reasonably assigned to the D–S stretching band of solid $D_2S$ (Fathe et al. 2006; Oba et al. 2019). The presence of $D_2S$ was further confirmed by detecting a desorption peak in the TPD experiment (Fig. 9b). The column density of $D_2S$ formed after 2 h of codeposition was estimated from the peak area and band strength (Fathe et al. 2006) to be approximately 3.5 ML, corresponding to a yield of approximately 15% (3.5/22.5). This was corroborated by the TPD–QMS experiment, which revealed desorption peaks at 80 and 155 K (m/z = 36). While other S-bearing species, such as DS(O)OD and DSOD, may contribute to the absorption bands observed at wavelengths less than 1100 cm$^{-1}$, we could not assign these peaks to specific molecules owing to the lack of literature data on their solid-state band positions and the unavailability of their chemical reagents.

Figure 9c shows TPD data at $m/z$ = 68 after the codeposition of SO$_2$ with D atoms or D$_2$ molecules (blank) for 2 h. In the blank experiment, a small desorption peak was observed at approximately 105 K, which was absent in the SO$_2$ + D codeposition experiment. Based on the desorption temperature and the peak area (~0.03% relative to SO$_2$ at $m/z$ = 64), this peak may be attributed to one of the SO$_2$ isotopologues, $^{34}$S$^{18}$OO, whose natural abundance is estimated to be approximately 0.01% (with $^{34}$S at 4.25% and $^{18}$O at 0.2%). Conversely, a new desorption peak observed at approximately 160 K may also have a contribution from $^{34}$S$^{18}$OO due to the formation of D$_2$O in the SO$_2$ + D experiment (Fig. 8). However, the observed peak area was about 0.7% of SO$_2$ at $m/z$ = 64 in the same experiment, which is significantly higher than the value of 0.03% in the blank experiment and the expected value of 0.01% based on the natural abundances of S isotopes as mentioned above. Hence, there should be larger contributions to the observed peak from other species such as DS(O)OD and/or DOSOD, which have a parent mass of 68.

In summary, our study on the chemical reactions of SO$_2$ with H atoms revealed the formation of S-bearing species such as HS(O)OH, HOSOH, and H$_2$S, as well as H$_2$O. When D atoms were used instead of H atoms, the corresponding D-substituted products were identified. The experimental results were consistent with the computational results (see section 5). The low yields of S-bearing products suggest that these species are partially lost from the substrate owing to chemical desorption upon formation. Although some infrared-inactive S-bearing species such as S atoms could form in the reaction sequence (e.g. H$_2$S + H → HS + H$_2$, HS + H → S + H$_2$), they will easily react with H atoms to yield HS and H$_2$S, which may be lost from the substrate upon the formation by chemical desorption (Oba et al. 2018). Hence, we expect that a contribution from these S-bearing species is not significant under the present experimental conditions.

## 5. COMPUTATIONAL RESULTS

*5.1 SO$_2$ + H reaction on ice*

The calculated binding energies of SO$_2$ on ice range from 0.37 to 0.58 eV, with an average binding energy of 0.45 eV (Fig. 10a). Given this relatively strong binding energy, SO$_2$ can effectively adhere to ice and react with H atoms on ASW. Figure 4 shows the calculated reaction pathways for the $SO_2$ + H reaction on ice. The initial addition of H to $SO_2$ (**1**) has a barrier of 0.07 eV, occurring at the S atom of $SO_2$ and producing the $HSO_2$ (**2**) radical with an exothermicity of –0.94 eV. In contrast, the barrier for H addition to an oxygen atom of $SO_2$ is 0.440 eV. Therefore, the initial H addition is more favorable at the S atom of $SO_2$. The second hydrogen addition to the (**2**) radical is barrierless, leading to the formation of either HS(O)OH (**3**) or $H_2SO_2$ (**3'**). Formation of (**3**) is more exothermic (−3.38 eV) compared to (**3'**) (–2.97 eV). Therefore, we conclude that the second hydrogen addition preferentially yields the thermodynamically stable product (**3**). Further hydrogenation of (**3**) can proceed via four possible reaction pathways. Among these, H atom abstraction from the HS– unit of (**3**) has the lowest reaction barrier (0.08 eV). The calculated reaction barriers for H addition to the S, OH, and O units of (**3**) are 0.23, 0.86, and 0.93 eV, respectively. While, the H atom abstraction from the -OH unit has a barrier of 0.68 eV. Consequently, the third hydrogen attack is favored at the H atom of the HS- unit of (**3**), resulting in the formation of the S(O)OH (**4**) radical. The subsequent barrierless fourth hydrogen addition to the (**4**) radical occurs preferentially at the OH unit, leading to the highly exothermic (–3.87 eV) formation of HO–S–OH (**5**). If the fourth hydrogen addition were to occur at the S or OH unit, the products (**5' = 3,** –3.72 eV) or SO···$OH_2$ (**5"**, –2.67 eV) could be formed. However, the exothermicity of forming (**5**) is greater than that of forming either (**5' = 3**) or (**5"**). Therefore, we conclude that the fourth hydrogen addition produces the thermodynamically most stable product (**5**). We then explored the hydrogenation pathways for (**5**). The addition of hydrogen to the S atom of (**5**) has a barrier of 0.13 eV, while the addition to one of the –OH units of (**5**) has a barrier of 0.69 eV. Thus, hydrogen addition to the S atom of (**5**) is favorable, leading to the formation of an HO···S(H)OH (**6**). Subsequent hydrogen addition to the OH radical in (**6**) results in the production of an $H_2O$ molecule and HSOH (**7**). We identified two reaction pathways for hydrogen addition to (**7**). The calculated barrier for hydrogen addition to the S atom of (**7**) is 0.09 eV, whereas the addition to the oxygen atom of (**7**) has a higher barrier of 0.55 eV. Consequently, hydrogenation of (**6**) yields $H_2S$ (**7**) and an OH radical. Finally, the hydrogenation of the OH radical produces an $H_2O$ molecule (**9**). Overall, the reaction between $SO_2$ and H produces $H_2S$, two $H_2O$ molecules, and $H_2$, following the overall reaction: $SO_2$ + 8H → $H_2S$ + $2H_2O$ + $H_2$. This process involves eight hydrogenation steps and results in the formation of several S-containing intermediates. The reaction pathway is as follows: $SO_2$ → $HSO_2$ (radical) → HS(O)OH → S(O)OH (radical) → HO–S–OH → HS–OH → $H_2S$.

## 6. ASTROCHEMICAL IMPLICATION

The presence of solid $SO_2$ in the ISM was first suggested in 1997 around the protostar W33A (Boogert et al. 1997). More recently, it has potentially been detected in dense molecular clouds by the JWST (McClure et al. 2023; Rocha et al. 2024). The observed solid $SO_2$ constitutes only 0.6–6% of the cosmic S abundance (Boogert et al. 1997) or approximately 0.4% (McClure et al. 2023), suggesting that other sulfur reservoirs may exist in space. This is consistent with our findings that solid $SO_2$ is readily consumed by reactions with H atoms on ice. If solid $SO_2$ forms from the condensation of gaseous $SO_2$, as predicted by some modeling studies (e.g., Laas & Caselli 2019), it is likely to be quickly consumed by reactions with H atoms. This would produce not only hydrogenated products such as HS(O)OH, S(O)OH, HOSOH, and HSOH but also other species such as $H_2S$ and $H_2O$. These processes are not accounted for in current astrochemical models (e.g., Vidal et al. 2017; Laas & Caselli 2019). However, as demonstrated in our $SO_2$ predeposition experiment, these species were not observed in the products (Fig. 2a). Nevertheless, these species can indeed be produced through a series of $SO_2$ hydrogenation reactions under certain conditions, as demonstrated in the codeposition experiment (Figs. 6 and 7). Some products are likely to desorb into the gas phase upon formation owing to chemical desorption (e.g., $H_2S$; Oba et al. 2018). For other products, such as $HSO_2$ and HS(O)OH, formed by reactions (5) and (6), respectively, their desorption probabilities were estimated using an empirical formula given by Garrod et al. (2007). This estimation incorporates the exothermicities of the reactions (0.94 and 3.38 eV) and the binding energies on ice (0.50–0.92 and 0.61–0.95 eV; Fig. 10). For $HSO_2$, the desorption probability is $6.6 \times 10^{-6}$, while for HS(O)OH, it is $9.5 \times 10^{-4}$. These values are considerably lower than those for other major interstellar species (see Wakelam et al. 2017). However, it is uncertain whether the empirical formula adequately explains the desorption probabilities (see discussion by Furuya et al. (2022)). Consequently, it is possible that such S-bearing species may effectively desorb from icy grains via chemical desorption. In either scenario, because S-bearing species formed through $SO_2 + H$ reactions may not have been targeted in astronomical observations or included in chemical models, they could be potential candidates for the "missing sulfur" in the ISM.

Conversely if $SO_2$ is formed from other S-bearing species embedded in ice through energetic processes (e.g., photolysis of $H_2S/H_2O$ ices; Jimenez-Escobar & Muñoz Caro 2011), it may also become incorporated in the bulk ice. In this case, solid $SO_2$ would have less opportunity to react with H atoms compared to reactions occurring on the ice surface, potentially leaving the solid $SO_2$ intact in the ice. Consequently, this form of solid $SO_2$ may be more readily detected by astronomical observations. This scenario aligns with the chemical modeling study by Shingledecker et al. (2020), which suggests that the primary source of $SO_2$ is in the bulk of dust-grain ice mantles after approximately $10^6$ years. The reported relative abundance of solid $SO_2$ in interstellar objects (~0.02%–1% compared to $H_2O$; Boogert et al. 1997; Zasowski et al. 2009; McClure et al. 2023; Rocha et al. 2024) and in cometary bodies (~1.2% compared to $H_2O$; Calmonte et al. 2016) suggests that $SO_2$ in bulk ice may be preferentially incorporated into planetary bodies. This incorporation could play an important role in further chemical evolution, such as the synthesis of S-bearing organic molecules, in these environments (Naraoka et al. 2022). Indeed, extensive analysis of samples returned from the carbonaceous asteroid Ryugu has identified several S-bearing species, including $S^{2-}$, $HSO_4^-$, $S_8$, and $CH_3SO_4^-$, in the aqueous extract (Yoshimura et al. 2023). This suggests a complex sulfur chemistry in the Solar System. A more detailed assessment of the distribution of S-bearing species both in the ISM and in the Solar System could provide a comprehensive understanding of sulfur's chemical evolution in space.

Acknowledgments

This work was partly supported by the JSPS KAKENHI grant numbers JP23H03980, JP21H04501 (to Y.O.), JP20H05847 (to K.F., N.W.), JP22H00159 (to N.W.), JP21F21319 (to T.N., N.W.). Supercomputing resources at the Institute for Information Management and Communication at Kyoto University in Japan and the Institute for Molecular Science (IMS) in Japan are acknowledged.

1717

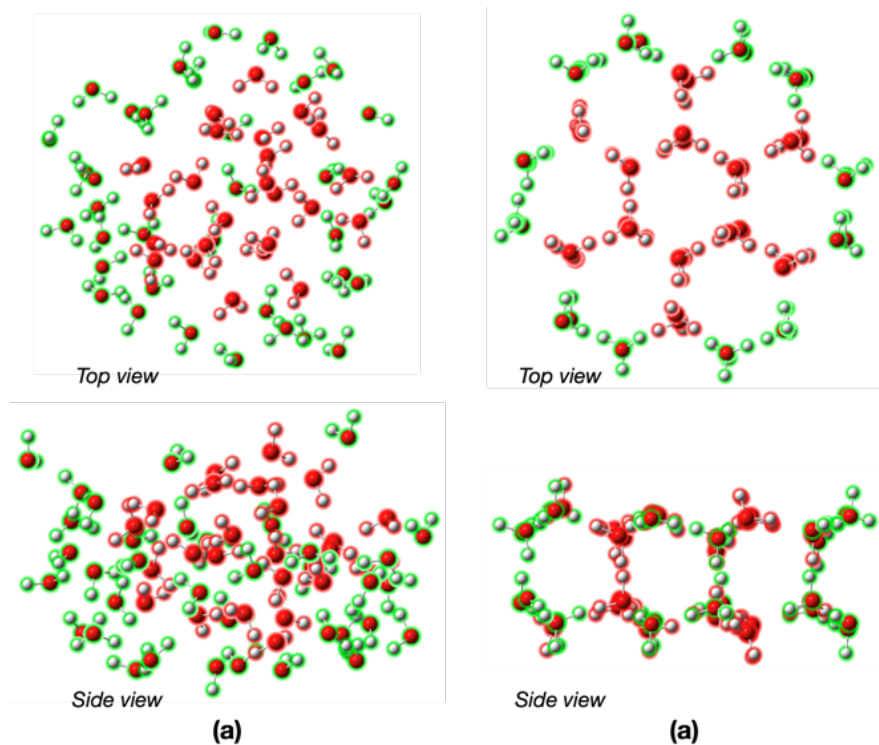

Figure 1 Top and side views of the ice cluster models (a) ASW and (b) Ih

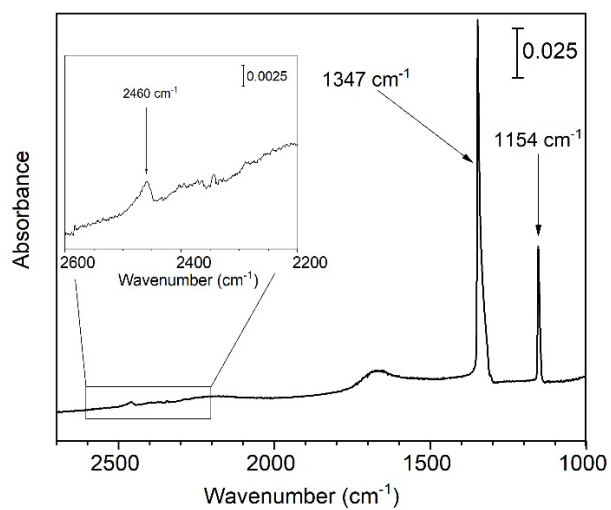

Figure 2 Fourier-transform infrared spectrum of solid $SO_2$ on c-ASW at 10 K. The inset provides a detailed view of the 2200–2600 cm$^{-1}$ region.

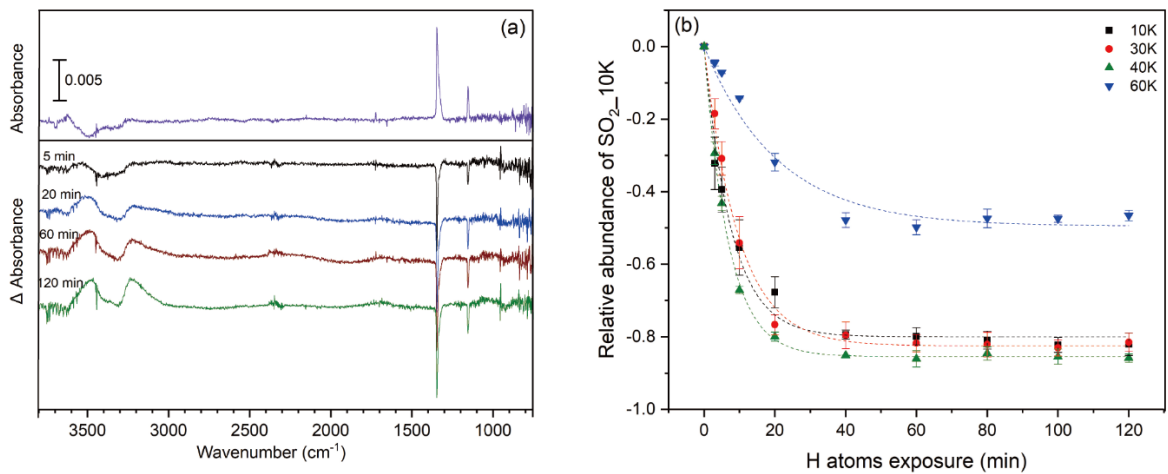

Figure 3 (a) Variation in FTIR spectra of $SO_2$ after exposure to H atoms on c-ASW for 5, 20, 60, and 120 min at 10 K (b) Relative abundance of $SO_2$ after exposure to H atoms for up to 2 h on c-ASW at 10, 30, 40, and 60 K.

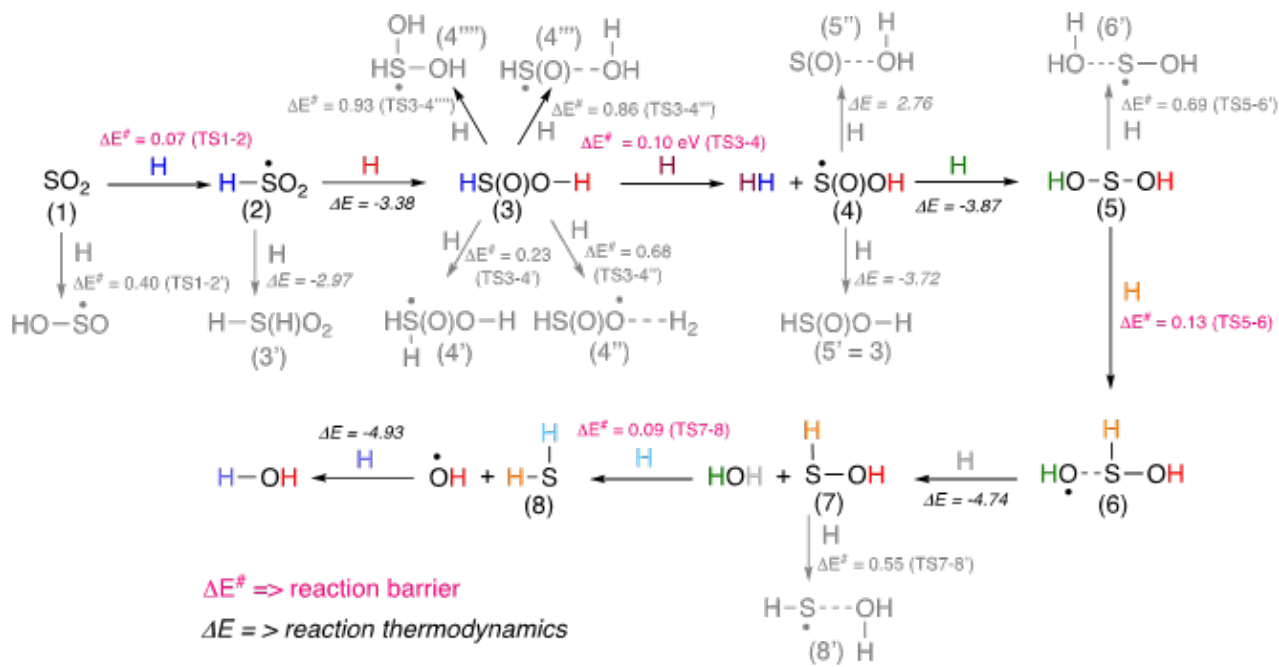

Figure 4 Calculated reaction pathways with relative energies expressed in eV. Cartesian coordinates of the optimized structures are in the Supporting Information (Table A1).

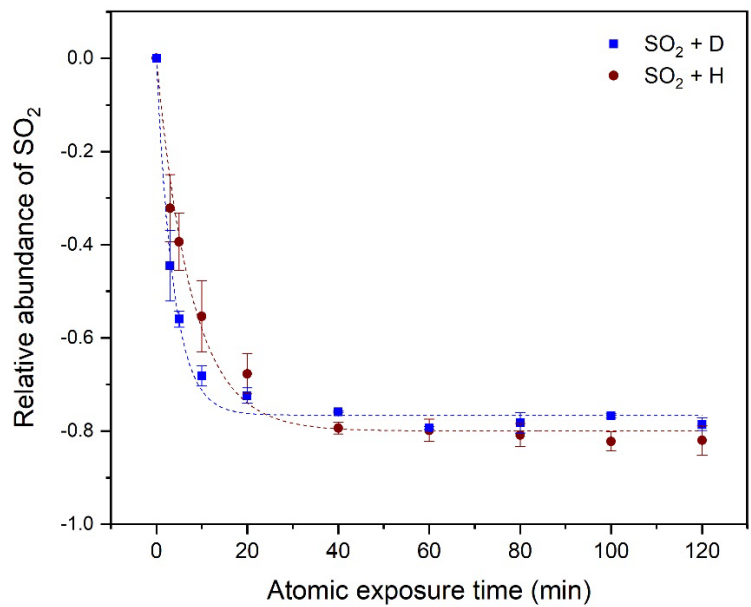

Figure 5 Changes in the relative abundances of solid $SO_2$ on c-ASW with exposure to H or D atoms at 10 K. Dotted lines represent fits to equation (15).

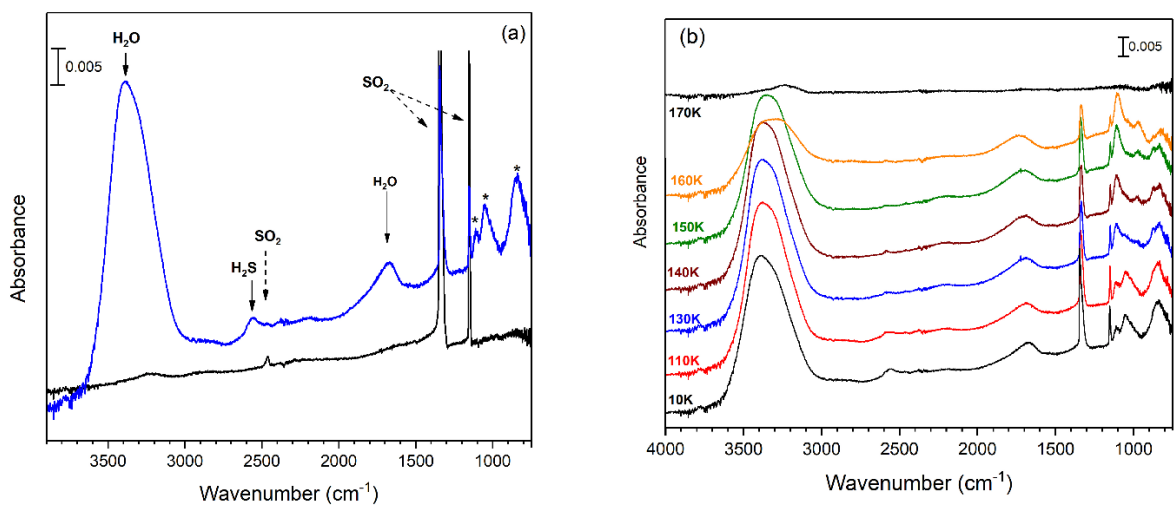

Figure 6 (a) Comparison of $SO_2$ after codeposition with $H_2$ (black line) and H atoms (blue line) for up to 2 h. Peaks at 1110 cm$^{-1}$, 1050 cm$^{-1}$, and 842 cm$^{-1}$ could be derived from S-bearing species such as $H_2SO$, HOSOH, and HS(O)OH. (b) Variations in the FTIR spectrum of products after the codeposition of $SO_2$ with H atoms as the temperature increases. Vertical dotted lines in (b) are guide for eyes.

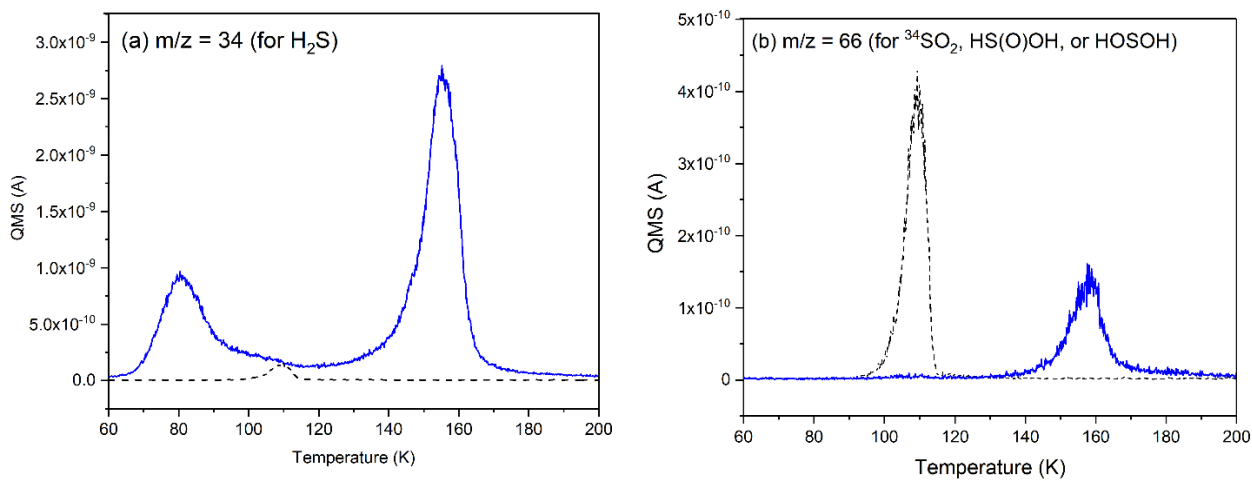

figure 7  TPD–QMS profiles after the codeposition of $SO_2$ with H atoms (solid line) or $H_2$ molecules (dotted line) at 10 K for up to 2 h: (a) at $m/z$ = 34 (for $H_2S$) and (b) at $m/z$ = 66 (for HS(O)OH). Peaks observed at ~80 and ~155 K correspond to the desorption of $H_2S$ from $H_2S$ and $H_2O$, respectively (Collings et al. 2004). A minor peak observed at ~110 K in panel (a) for the blank experiment is attributed to the fragment ($^{34}S$) of $^{34}SO_2$.

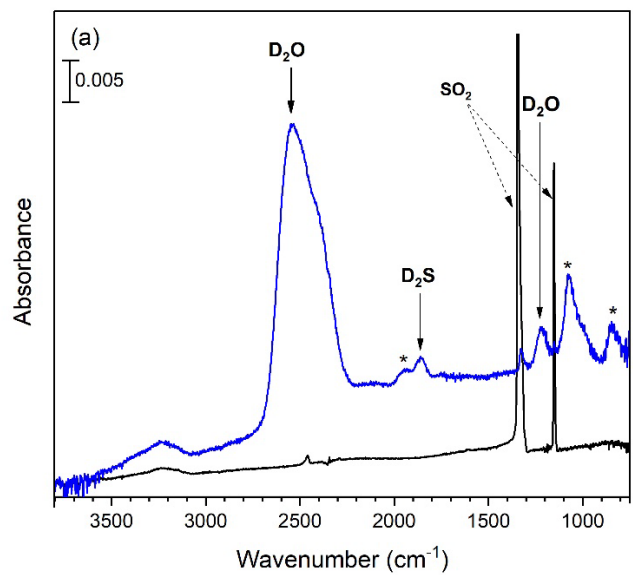

Figure 8  FTIR spectrum of the codeposition of $SO_2$ with D atoms (blue) compared to $D_2$ molecules as a blank (black) for 2 h at 10 K. Peaks at 1948 cm$^{-1}$, 1075 cm$^{-1}$, and 847 cm$^{-1}$ could be derived from S-bearing species such as $D_2SO$, DOSOD, and DS(O)OH.

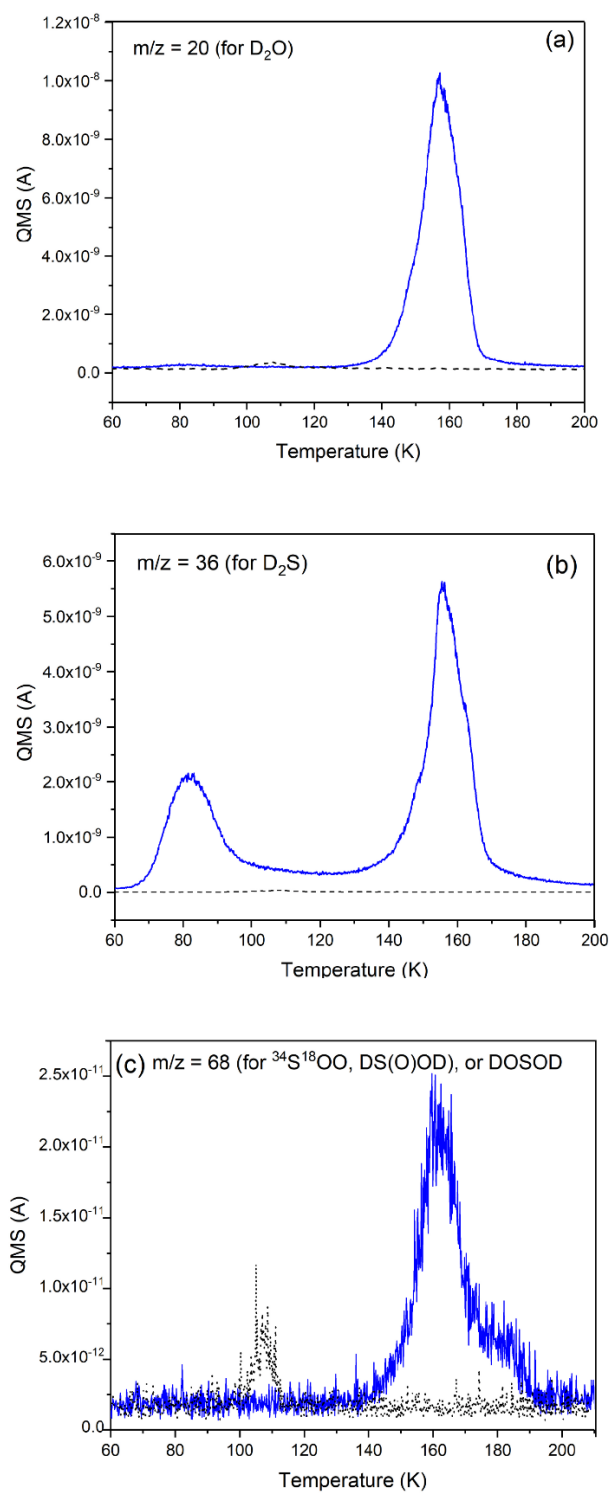

Figure 9  TPD–QMS profiles after the codeposition of $SO_2$ and D atoms (solid line) and $D_2$ molecules (dotted line) for 2 h at 10 K: (a) $m/z = 20$, (b) $m/z = 36$, and (c) $m/z = 68$

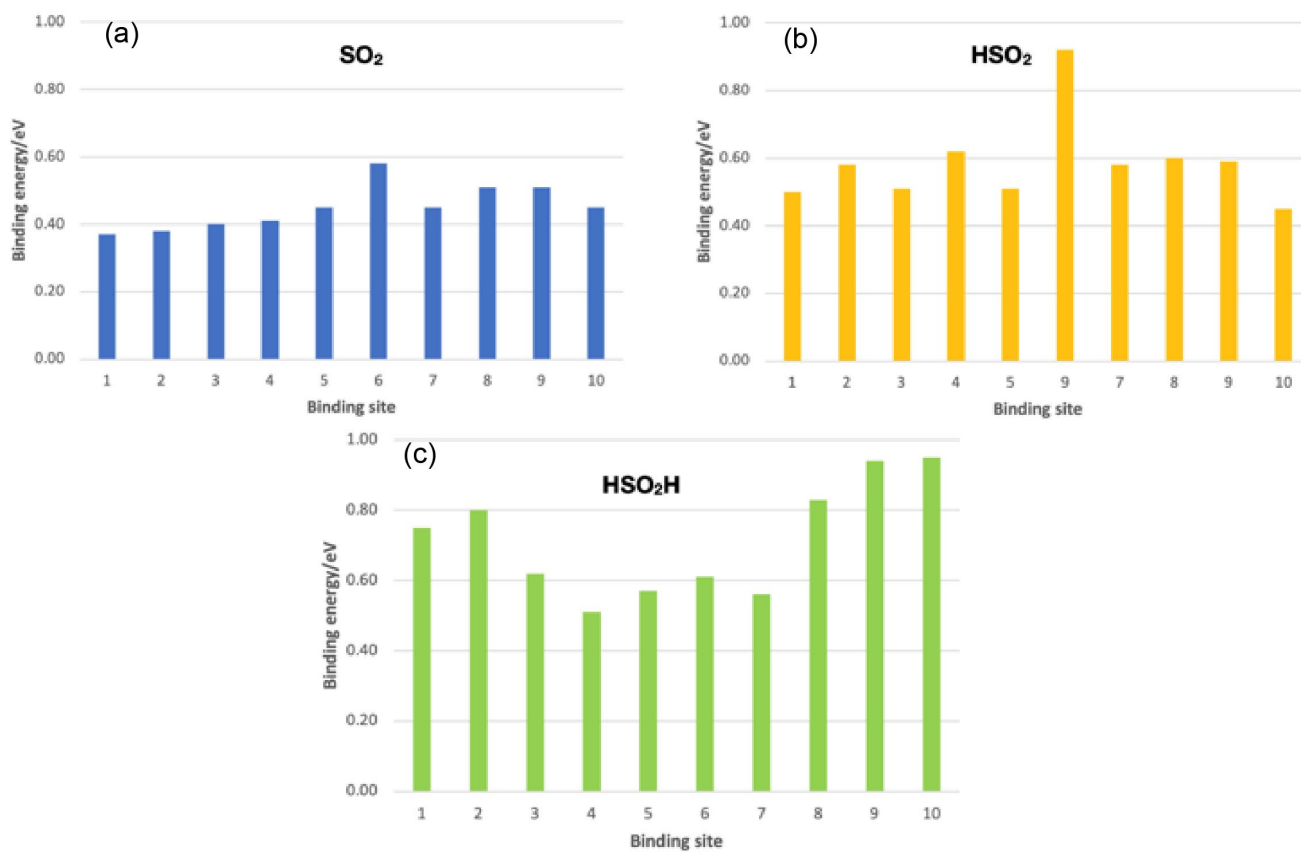

Figure 10  Binding energies of (a) SO$_2$, (b) HSO$_2$, and (c) HSO$_2$H on ASW (1–5) and I$_h$ (6–10)

**APPENDIX**
Cartesian coordinates of the optimized structures are in Table A1.

**Table A1. Cartesian coordinates of the optimized local minima and transition states.**

| 1 | X Cartesian Coodinate | Y Cartesian Coordinate | Z Cartesian Coordinate |
|---|---|---|---|
| S | 5.65968200 | 2.97243000 | 0.32039600 |
| O | 4.61028100 | 3.38888000 | 1.20012000 |
| O | 6.90729800 | 2.58516700 | 0.89571700 |
| H | 6.31566200 | -0.70790300 | -0.35781200 |
| O | 7.92667400 | -3.28595800 | -3.71700000 |
| H | 8.76938600 | -3.10090900 | -4.16926400 |
| H | 7.34619300 | -2.51470000 | -3.94152000 |
| O | 6.03980300 | 9.55674600 | -7.53374500 |
| H | 5.37091600 | 8.86068400 | -7.35846800 |

Only a portion of this table is shown here to demonstrate its form and content. A machine-readable version of the full table is available.